\documentclass[aps,pra,twocolumn,showpacs]{revtex4}
\usepackage{graphicx}
\usepackage{bm}
\usepackage{amsmath}
\usepackage{latexsym}
\usepackage{float}
\newfloat{widefig}{thp}{lop}
\begin{document}
\title{Correlation studies in a spin imbalanced fermionic superfluid in presence of a harmonic trap }
\author{Poulumi Dey} %
\email{poulumi@iitg.ernet.in}%
\author{Saurabh Basu}
\email{saurabh@iitg.ernet.in}%
\affiliation{Department of Physics, Indian Institute of Technology
Guwahati, Guwahati, Assam 781039, India } 
\author{R. Kishore}%
\email{kishore@las.inpe.br}%
\affiliation{Instituto Nacional de Pesquisas Espaciais, CP 515, S.J. Campos, SP, 12245-970, Brazil}%

\date{\today}

\begin{abstract}

We investigate different physical properties of a spin imbalanced fermionic superfluid described by an attractive 
Hubbard model in presence of a harmonic trap. To characterize the ground state of
such a system, we compute various correlation functions, such as pairing, density and current correlations
To underscore the effect of the trap, we compute these quantities as a function of strength 
of the trapping potential. A distinguishing feature of a superfluid state with population
imbalance, is  the appearance of a phase with spatially modulating gap function, 
called the Fulde-Ferrell-Larkin-Ovchinnikov (FFLO) phase which is present for both with and without 
trap. Trapping effects induce lowering of particle density and yields an {\it{empty}}
system in the limit of large trapping depths.

\end{abstract}

\maketitle

\section{Introduction}

The experimental realization of Bose-Einstein condensation in dilute alkali
 gases\cite{Anderson,Han,Ernst,Esslinger,Davis} has triggered vast exploration 
of the field of ultracold, 
degenerate gas of fermionic atoms in presence of a harmonic trap. Achieving degeneracy in a system of fermionic 
atoms is a well known hurdle as $s$-wave collisions between fermions in the same state are blocked by the 
exclusion principle (scattering in higher angular momentum channels, such as $p$-wave, are anyway
weak). However researchers continued to access the degeneracy limit and the rare earths are recent additions on the list\cite{McNamara,Fukuhara}. In most atomic gas experiments, the experimental challenge of producing ultracold temperatures
is achieved by confining the atoms using a trapping potential. In fact, the velocity or momentum distribution
of the gas is obtained via a technique called 'time of flight' which involves a {\it{sudden turn off}} 
of the trapping potential and hence wait for an {\it{expansion time}}. The trapping potential 
is produced due to interaction between a gaussian laser beam and the induced atomic dipole moment.
At the bottom of the trap, the gaussian profile can pretty much be approximated by a harmonic function,
except for very shallow trap depths.

In this paper, we shall incorporate the optical trap depth effects by a harmonic trap. So our system comprises of a 
population imbalanced superfluid in the presence of a harmonic trap which has a minimum (deepest) at the center
of the (two-dimensional) lattice and gradually increases towards the edges\cite{Chen,Iskin}. Further we shall focus on the effect 
of the trap on properties of the ground state by reviewing different correlation functions that
characterize the state. These are sure to provide important insights on the nature of the 
polarized Fermi gas.

A convenient technique to emphasize the effect of the trap is to examine the scenario in absence of a trap
and compare it with those for various trap depths. Further, since the presence of a harmonic trap
forces the system, which is a two-dimensional superfluid of Fermi atoms, to loose translational invariance, it
is prudent to investigate correlations in the real space. In this paper, we include several of them
such as, gap parameter, off-diagonal long range order, pair-pair, density-density and current-current
correlations to study the ground state of the population imbalanced Fermi gas in presence of a harmonic trap.
The above list may not be exhaustive, but are certainly useful to characterize a state that
incorporates physics of superfluidity in presence of combined effects of population imbalance
and harmonic trap.

The results that we obtain have two special features. A phase with a finite momentum Cooper pairing 
and characterized by a spatially modulating order parameter, exists for moderate 
values of polarization of the participating species of fermions. 
This state, called as the Fulde-Ferrell-Larkin-Ovchinnikov (FFLO) state,
persists for both with and without trap\cite{Loh}. However large imbalance gives way to a fully polarized
Fermi liquid\cite{Chandrasekhar}. Another interesting artifact of the presence of the trap occurs 
 in the form of loss in density of the constituent particles of the system. This
is inevitable since the chemical potential is required to be kept fixed to attain
a population imbalance of the participating species and thus is crucial for observing
the FFLO phase. Hence in the limit of large trap depth, we obtain an almost {\it{empty}}
system. This almost {\it{empty}} system has some features that mimic an insulating 
phase, such as a vanishing superconducting gap, an increasing spectral gap and a 
dramatic reduction in mobility of the particles as a function of the trapping potential.
It should be noted that a gradual emergence of such an {\it{empty}} system with increasing 
trapping effects {\it{can not}} be termed as a transition to an insulating state which are 
otherwise present in the repulsive version of the Hubbard model\cite{Jordens}.
 Such transition have been experimentally realized in case of bosonic 
atoms\cite{Greiner} that are supported by theoretical studies\cite{Zwerger}.

The study of population imbalanced gases of fermionic atoms in presence of a harmonic trap has a reasonable
history\cite{Iskin,Chen,Stoferle,Strohamair,Koetsier}. Density distribution of the fermionic atoms 
in the trap has been a central issue and a phase separation is predicted with the paired atoms confined
at the center of the trap, while the unpaired ones find place towards the periphery\cite{Liu1,Liu2}.
However a concomitant tuning of the spin imbalance and trap depth was lacking. This paper fills the void
and explores the interesting physics associated with it.

We organize our paper in the following fashion. In order to fix notations, we provide a description of the
attractive Hubbard model in presence of a harmonic trap and the standard mean field Bogoliubov de Gennes (BdG) 
theory which are used here. In the next section, we present 
results for different correlation functions in the presence of a trap and compare with the results when 
the trapping effects are switched off. For a range of population imbalance caused by a constant Zeeman field in our 
model, signature of FFLO phase is observed. The FFLO phase is characterized by weaker correlations
but has a robust presence both with and without trap effects\cite{Chen}. Further as a function 
of the trap depth, we demonstrate a gradual lowering of the particle density in table I. A probe 
into the nature of the almost {\it{empty}} state yields a zero superconducting gap, an increasing
spectral gap and a reduced average kinetic energy of the particles as the strength of
confinement is enhanced. These features may indicate emergence of an insulating-like state,
however when considered along with a gradual vanishing of particle density, it rules 
out the possibility of a transition from a superfluid to an insulator. We conclude with a brief discussion
of our results.

\section{Model and Formalism} 

In crystal lattices, where electrons are the main carriers that contribute to transport properties of materials,
obtaining a model Hamiltonian that describes the physical properties satisfactorily is a difficult task
owing to a highly complicated band structure. However a gaseous system of fermionic atoms in a confining
potential is a much neater realization of the simplest model that incorporates strong correlations 
between atoms at short distances.  An attractive version of the two-dimensional Hubbard model 
in which the atoms interact attractively via the on-site interaction,
$U$ in presence of a constant Zeeman field, $h$ and a harmonic trapping potential at site $r_{i}$, $V_{i}$ is written as,

\begin{eqnarray}
\label{eq:hubbard_model}
{\cal{H}} = &  -t &  \sum_{\langle ij \rangle,\sigma}(c^{\dagger}_{i\sigma}
c_{j\sigma} + {\rm {H.c.}}) \\ \nonumber
 & - &  \left|U\right| \sum_{i}(n_{i\uparrow}-\frac{1}{2}) (n_{i\downarrow}-\frac{1}{2})
  +  \sum_{i,\sigma}(V_{i}-\mu+\sigma h)n_{i\sigma}
\end{eqnarray}
$\mu$ denotes the chemical potential, $t$ is the hopping matrix element between the nearest 
neighbours of a two dimensional square lattice. The creation (annihilation) operator for fermionic atoms corresponding
to spin $\sigma$ is $c^{\dagger}_{i\sigma}$ ($c_{i\sigma}$). The excess of one
species of atoms (say with spin-$\uparrow$) over another is controlled by the magnetic field
$h$ (or equivalently an effective chemical potential, $\mu' = \sigma h - \mu$). $V_{i}$ is 
assumed to be of the form,
\begin{equation}
V_{i} = V_{0}(r_{i}-r_{0})^{2}
\end{equation}
where $V_{0}$ is the strength of the trapping potential and $r_{0}$ is the position where the
center of the trap lies and is located at the center of the lattice. Thus the potential is minimum
(deepest) at the center of the lattice and is maximum (shallow) at the edges.
All of $U$, $h$, $\mu$ and $V_{0}$ are
expressed in units of hopping strength, $t$ which is typically of the order of an eV.
A mean field decoupling of the interaction term in
 Eq.~(\ref{eq:hubbard_model})   yields the effective Hamiltonian of the form,
\begin{equation}
\label{eq:eff_model}
{\cal{H}}_{eff}  =   \sum_{ij,\sigma}{\cal{H}}_{ij\sigma}
(c^{\dagger}_{i\sigma}c_{j\sigma}
+ {\rm {H.c.}}) 
 +   \sum_{i}\left[\Delta_{i} 
c^{\dagger}_{i\uparrow}c^{\dagger}_{i\downarrow}- \Delta^{*}_{i}c_{i\uparrow}c_{i\downarrow}\right]
\end{equation}
here $\Delta_{i} = -|U|\langle c_{i\downarrow}c_{i\uparrow}\rangle$
is the gap parameter for the fermionic superfluid.  
${\cal{H}}_{ij\sigma}=-t \delta_{i\pm 1j} + (V_{i} - \mu - U \delta n_{i\bar{\sigma}} +\sigma h) \delta_{ij}$ 
where $\delta n_{i\bar{\sigma}}=n_{i\bar{\sigma}}-1/2$ with
 $\langle n_{i\sigma}\rangle = \langle c^{\dagger}_{i\sigma}c_{i\sigma}\rangle$
and $\bar{\sigma} = -\sigma$.  

Eq.~(\ref{eq:eff_model}) is hence diagonalized using Bogoliubov transformation which yields,
\begin{eqnarray}
\label{eq:matrix}
\left(\begin{array}{cc}
{\cal{H}}_{ij\sigma} & \hat{\Delta_{i}}\\
\hat{\Delta_{i}}^{*} & -{{\cal{H}}_{ij\bar{\sigma}}^{*}}\end{array} \right)\left( \begin{array}{c}
u_{n}({\bf {r}}_{i})\\
v_{n}({\bf {r}}_{i})\end{array} \right)=E_{n} \left( \begin{array}{c}
u_{n}({\bf {r}}_{i})\\
v_{n}({\bf {r}}_{i})\end{array} \right)
\end{eqnarray}
where 
$u_{n}({\bf {r}}_{i})$ and $v_{n}({\bf {r}}_{i})$ are the BdG eigenvectors
satisfying $\sum_{n}[u^{2}_{n}({\bf {r}}_{i}) + v^{2}_{n}({\bf {r}}_{i})] = 1$ 
for all ${\bf {r}}_{i}$ and $E_{n}$ are the eigenvalues.

The gap parameter and density (and hence magnetization, 
$m_{i} = \langle n_{i\uparrow}\rangle - \langle n_{i\downarrow}\rangle$, see Eq.~(\ref{eq:selfcon2})) in 
terms of the eigenvectors $u_{n}({\bf {r}}_{i})$ and $v_{n}({\bf {r}}_{i})$ at 
 a temperature $T$ are given by,
\begin{eqnarray}
\label{eq:selfcon1}
\Delta_{i}   =   & - & |U| \sum_{n} [u_{n}({\bf {r}}_{i})v^{*}_{n}({\bf {r}}_{i})f(E_{n\uparrow})  \\ \nonumber
&  - &   u_{n}({\bf {r}}_{i})v^{*}_{n}({\bf {r}}_{i}) f(-E_{n\downarrow}) ] 
\end{eqnarray}
\begin{equation}
\label{eq:selfcon2}
\langle n_{i\sigma}\rangle  = \sum_{n} \left[|u_{n}({\bf {r}}_{i})|^{2}f(E_{n\sigma}) + |v_{n}({\bf {r}}_{i})|^{2}f(-E_{n\bar{\sigma}}) \right] 
\end{equation}
where  $f(E_{n\sigma})$ is the  Fermi distribution function.
$\Delta_{i}$ and $\langle n_{i\sigma}\rangle$ are obtained 
self-consistently from Eq.~(\ref{eq:selfcon1}) and Eq.~(\ref{eq:selfcon2})
at each lattice site.

It may be noted that a number of self-consistent solutions may exist for the 
gap parameter, $\Delta_{i}$ corresponding to one set 
of parameters with different initial guesses. The winner among these will be decided by computing the
free energies, $\cal{F}$ computed with respect to the free energy of vacuum
(at zero temperature) and is given by\cite{Qinghong},
\begin{eqnarray}
\label{eq:gse}
{\cal{F}} & = &  \sum_{n\sigma} E_{n\sigma} \left[f(E_{n\sigma})-\sum_{i} |v_{n}({\bf {r}}_{i})|^{2} \right] \\ \nonumber
  & + &  |U| \sum_{i}\langle n_{i\uparrow} \rangle \langle n_{i\downarrow} \rangle 
 +  \frac{1}{|U|} \sum_{i} \Delta_{i}^{2} - \frac{|U|N}{4}
\end{eqnarray}

We further compute higher order  correlations  which are likely to be more illustrative in characterizing
the phase. We begin with off-diagonal long range order (ODLRO) which characterizes the superfluid 
phase and can be useful in bringing out distinction between a zero net center of mass
momentum pairing (BCS) and a finite momentum pairing states (FFLO). 
The ODLRO order parameter, $\Delta_{op}$
is defined  by the long distance behaviour of the  correlation 
$\langle c^{\dagger}_{i\uparrow}c^{\dagger}_{i\downarrow}c_{j\downarrow}c_{j\uparrow} 
\rangle\rightarrow \Delta_{OP}^{2}/|U|^{2}$ for  
$|{\bf {r}}_{i}-{\bf {r}}_{j}|\rightarrow \infty$\cite{Waldram}. The other
correlation functions which are of physical significance are the pair-pair 
($ C_{ij} = \langle c^{\dagger}_{i\uparrow}c^{\dagger}_{i\downarrow}c^{\dagger}_{j\uparrow}c^{\dagger}_{j\downarrow}\rangle$), 
density-density ($ K^{\sigma \sigma^{'}}_{ij} = \langle c^{\dagger}_{i\sigma}c_{i\sigma}c^{\dagger}_{j\sigma^{'}}c_{j\sigma^{'}} \rangle$)
and paramagnetic current-current correlation  functions,
 $\Lambda^{\alpha\beta}_{ij} = \langle {j_{\alpha}}(r_{i}){j_{\beta}}(r_{j}) \rangle $ 
where ${j_{\alpha}}$ is the component of the paramagnetic current density operator at site $r_{i}$ given by 
${j_{\alpha}}(r_{i}) = it \sum_{\sigma} \left [ c^{\dagger}_{i+\alpha,\sigma}c_{i,\sigma} - 
c^{\dagger}_{i,\sigma}c_{i+\alpha,\sigma} \right ]$. Here $r_{i}$ and $r_{j}$ denote lattice
sites. In our numerical computation, a pair (say for $C_{ij}$) is located at a fixed 
site $r_{i}$ and the probability of another pair at site $r_{j}$, where $r_{j}$ can be
any other site of the lattice, is calculated. Similar notation is followed for  $ K^{\sigma \sigma^{'}}_{ij} $ and $\Lambda^{\alpha\beta}_{ij}$ as well. It is useful to mention here that
while we have considered site $r_{j}$ to be along any arbitrary direction with respect to
site $r_{i}$, however results (shown in Fig.~6-8) are presented for $r_{j}$ to be along
the length ($x$-axis) of the lattice. The reason is the following. While the direction along which 
$r_{j}$ is chosen is immaterial for the homogeneous BCS phase, for the FFLO phase,
the preferred direction (for the case of without trap) is set by the propagation of the 
modulating gap parameter, which is along the length ($x$-axis) of the lattice (Fig.~2(c)). To retain
uniformity, we follow the same for the trapped case, where modulation exists
along a ring away from the trap center.

Finally, we comment on the choice of parameters. We perform our studies for an inter-particle
attraction strength $|U|=2.5t$ which may be considered to be weak, and hence
suitable as we propose a BCS superfluid as a starting point.
It should be noted that further lower values of $|U|$ yield a vanishing gap parameter 
for any reasonable value of density.
We have chosen $\mu$ to be $-0.5t$.
 All quantities are calculated at zero temperature, where the 
Fermi function has been taken as unity. We have also considered other values for density, ranging from
moderate to large (very small densities do not yield a stable superfluid phase) and 
different (weak) interaction strengths, $U$, however we have not noted
any qualitative change in behaviour of the correlation functions which are on focus in this paper. 
Further, it may be noted that the gap parameter undergoes a one
dimensional modulation\cite{Wang} with a period that
 is commensurate with the lattice size in the absence of confining potential. 
Thus  a  two dimensional lattice of size $32 \times 16$ 
has been considered so that more periods of modulation of the 
gap parameter can be accommodated. We have also checked that the results for a $32 \times 32$ lattice are 
unaltered with regard to the modulation of the gap parameter and qualitative behaviour of the
correlation functions. In case of presence of the external harmonic potential, 
the dimension of the lattice is chosen to be $24 \times 24$ with open boundary conditions 
and the trap center is located at the center of the lattice.

\section{Results} 

To characterize the spin polarized phase in presence of a harmonic trap, we have computed various
 correlation functions in the real space for various values of population imbalance and trap depth. 
The different correlations that we consider in this work are gap parameter, magnetization,  off-diagonal  long range order (ODLRO), pair, density and current correlations. 
The list is not exhaustive but will serve our purpose
in providing insights on the nature of the population imbalanced superfluid in the superposed 
environment of a harmonic trap.
 
Before we proceed with the discussion on the results for various correlation functions, we 
wish to distinguish between the three phases, {\it{viz.}} the homogeneous BCS phase for low values of
magnetic field ({\it{i.e.}} low population imbalance), a spatially inhomogeneous FFLO phase
for moderate field strengths and finally a {\it{normal}} phase at higher magnetic fields.
Thus to fix our input parameters, we  present a schematic diagram of the BCS and FFLO phases, 
both in the absence and presence of trap
in Fig.~1. The FFLO phase is marked by a lower and a upper critical magnetic field values,
 $h_{c_{1}}$ and $h_{c_{2}}$ respectively and lie intermediate to the BCS 
and normal phases. These boundaries are obtained from the behaviour of the gap parameter
 obtained solving Eq.~(\ref{eq:selfcon1}). Fig.~1 shows
 that the values for $h_{c_{1}}$ and $h_{c_{2}}$  are $0.35t$ and $0.5t$  in 
the absence of trap whereas it shifts to $h_{c_{1}}=0.25t$ and $h_{c_{2}}=0.4t$ when the trapping potential is switched on
 for $|U|=2.5t$ and $\mu=-0.5t$. Hence all our results are presented for 
two values of magnetic field, $h$ which denote
two different values of population imbalance.
One being in the homogeneous BCS phase ($h < h_{c_{1}}$) and another in the FFLO phase ($h_{c_{1}} < h < h_{c_{2}}$).
From Fig.~1, we choose $h=0.1t$ and $h=0.5t$ corresponding to no trap and $h=0$ and $h=0.3t$ in
presence of the harmonic trap for BCS and FFLO phases respectively. Moreover, we have restricted 
the demonstration and discussion of the results to one value of strength of the trapping potential
{\it{viz.}} $V_{0}=0.016t$. 
The representative value considered here, approximately denotes a point where there possibly
exists a very shallow minimum in the spectral gap.  We have investigated the correlations for other values of trapping 
depths in both the superfluid (small $V_{0}$) and strongly confined regimes 
($V_{0} \sim 0.2t$, see Figs.~ 5 and 9 and related discussion).

\begin{figure}[!htb]
\includegraphics[width=0.38\textwidth,height=1.8in]{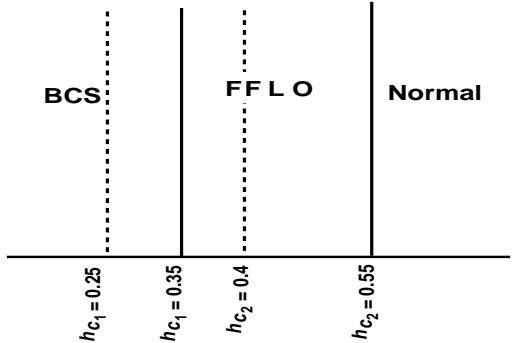}
\caption{A schematic representation of the FFLO phase is shown as a function of
the magnetic field both in the absence and presence of trap for $|U|=2.5t$ and $\mu=-0.5t$
 (same for all plots).
 This phase is intermediate to a BCS phase (homogeneous $\Delta_{i}$)
and a normal phase ($\Delta_{i}=0$). The boundaries of the FFLO phase are marked
by $h_{c_{1}}$ and $h_{c_{2}}$. The solid lines are the boundaries in the absence of trap
 whereas the broken lines mark the boundary in the presence of the trapping potential of strength,
 $V_{0}=0.016t$. Same numerical value of $V_{0}$ is chosen for Figs.~2-4 and Figs.~6-8.
 Also all the quantities  are in units of hopping integral, $t$.} 
\end{figure}

We now present our results for the  gap parameter, $\Delta_{i}$ in presence of a 
harmonic confinement and compare  with the corresponding behaviour  in  absence of the trap.
 It may be noted that  $\Delta_{i}$ is homogeneous for low values of magnetic 
 fields (population imbalance being small) in the absence of trap, whereas 
the trapping effects lead to a  maximum of $\Delta_{i}$ at the trap 
center, hence decreases and finally vanishes away from the center (see Fig.~ 2 (a) and (b)).
Such contrasting behaviour is caused by the trapping effects of the harmonic potential that results
in the accumulation of atomic densities around the center thereby leading to 
such an inhomogeneous distribution of $\Delta_{i}$ in real space.
Next we compare $\Delta_{i}$ for the modulating (FFLO) phase. $\Delta_{i}$ modulates
  along one direction ($x$-axis)
 of the lattice without the trap while it modulates along the radial  direction
in the presence of trap, as shown in Fig.~2 (c) and (d). That is, far away from the trap center, the
gap parameter undergoes a sign change confirming the presence of the FFLO phase.
 
\begin{figure*}[!htb]
\includegraphics[width=1.0\textwidth,height=5.5in]{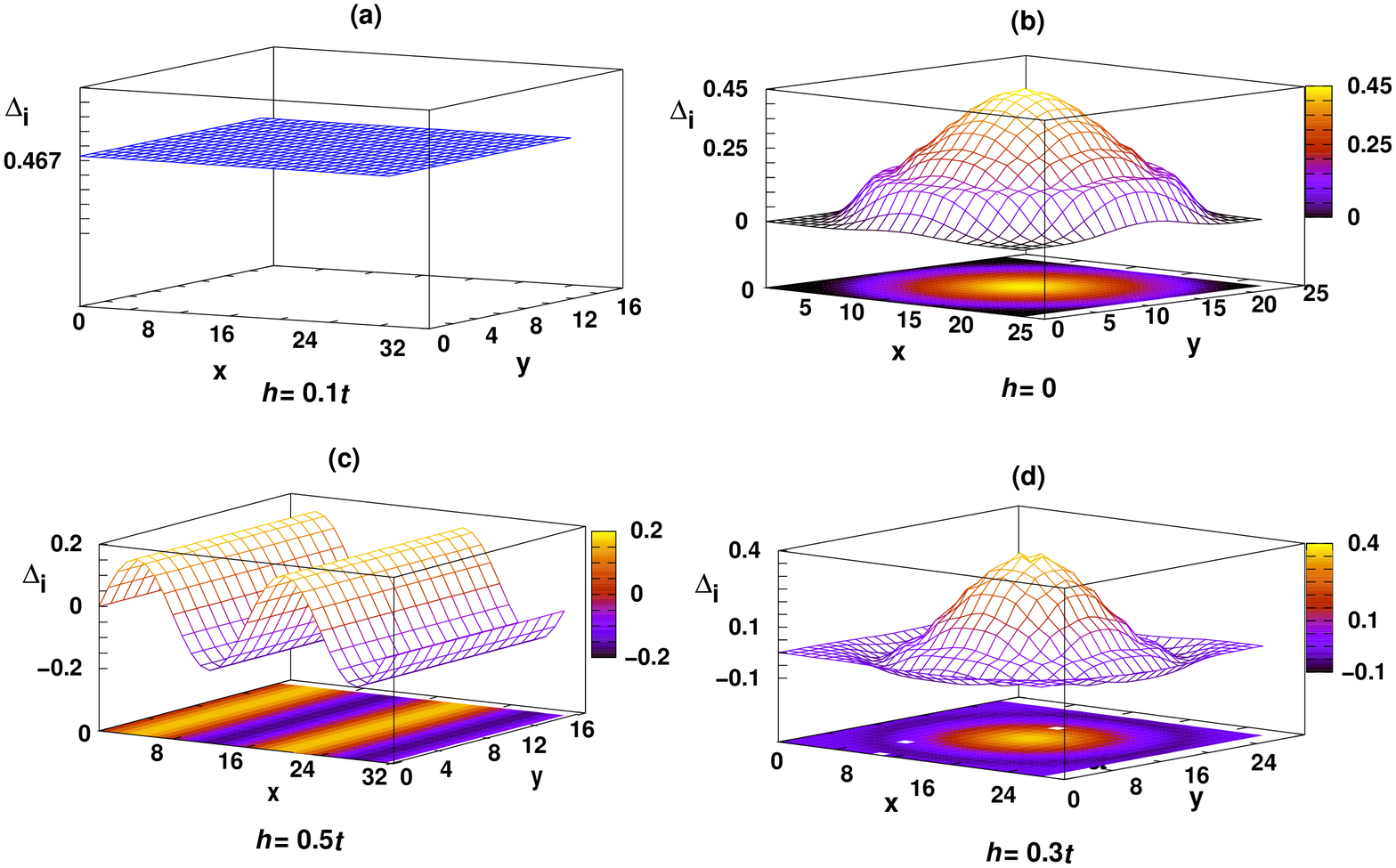}
\caption{(Colour online) Local pairing amplitude, $\Delta_{i}$ in BCS ((a) and (b))
and  modulating FFLO ((c) and (d)) phases are shown.  
 Figs. (a) and (c) are for cases without trap while (b) and (d) are in presence of trap. 
The system size is $32 \times 16$ in the absence of trap and $24 \times 24$ when 
the trapping potential is present and are same for all our results.} 
\end{figure*}

We proceed to discuss results on the local magnetization, $m_{i}$ ($= \langle n_{i\uparrow} \rangle - \langle n_{i\downarrow} \rangle$) 
presented in Fig.~3. While $m_{i}$ for the  BCS phase, is zero understandably (being non-magnetic due
to lack of unpaired particles) across the lattice (see Fig.~3(a) and
 (b)), it  modulates with the period  half of that of $\Delta_{i}$ in the FFLO phase when
trapping effects are not included (Fig.~3(c)). The large values of magnetization occurs at nodal lines with  broken pairs, 
whereas lattice sites with finite $\Delta_{i}$ correspond to weak magnetization, thereby leading to a phase 
difference between these two quantities. The magnetization profile in the presence of trap, exhibits a bimodal structure 
(as shown in Fig.~3(d)) since the magnetization is non-zero  around the ring like nodal line. In other
words, a small number of particles are squeezed into the inner core of the harmonic trap due to pair formation 
in the superfluid phase, while the unpaired (majority) carriers are pushed out to outside of the core.

\begin{figure*}[!htb]
\includegraphics[width=1.0\textwidth,height=5.5in]{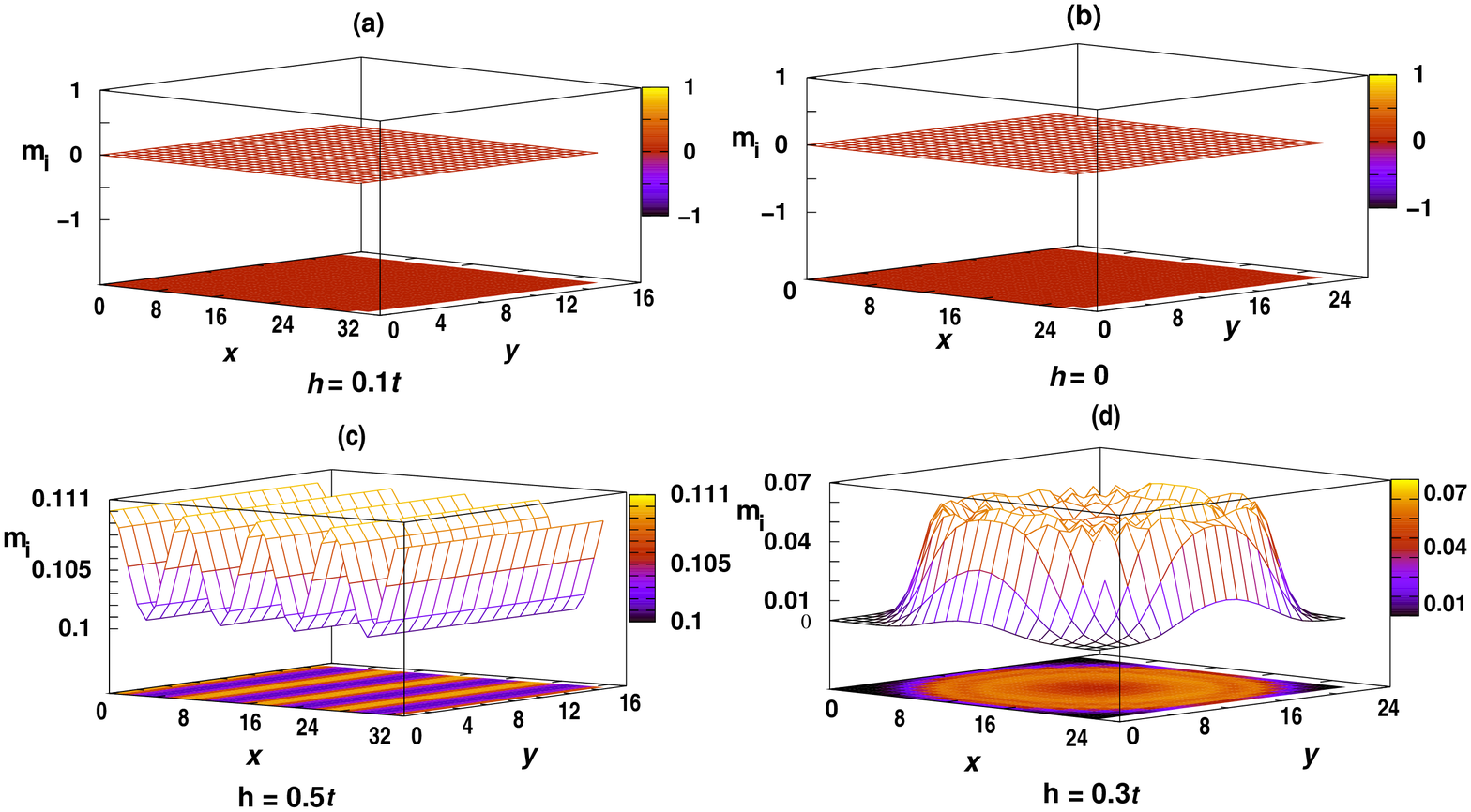}
\caption{(Colour online) Local magnetization, $m_{i}$ is shown
 in BCS (a) and (b) and FFLO (c) and (d) phases both in the absence and presence of the 
confining potential.} 
\end{figure*}

We now plot the ODLRO order parameter, $\Delta_{op}$ 
as a function of magnetic field  without the harmonic trapping in Fig.~4. It shows sharp drop
 of $\Delta_{op}$ at the onset of the FFLO phase at $h_{c_{1}}=0.35t$, thereby indicating  weakened
 superconducting correlations in the FFLO phase (see Fig.~1).  $\Delta_{op}$ is also computed in the
 presence of trap but 
it is seen at two fixed values of magnetic field {\it{viz.}} $h=0$ and $h = 0.3t$ for various values of trapping potential, $V_{0}$ 
(shown in  Fig.~5). Our results clearly indicate a higher value of $\Delta_{op}$
   for $h=0$ (no population imbalance) than that for $h=0.3t$, but is suggestive 
of a rapid decay for both with  increasing confinement of the carriers.
 
 \begin{figure}[!htb]
\includegraphics[width=0.4\textwidth,height=1.8in]{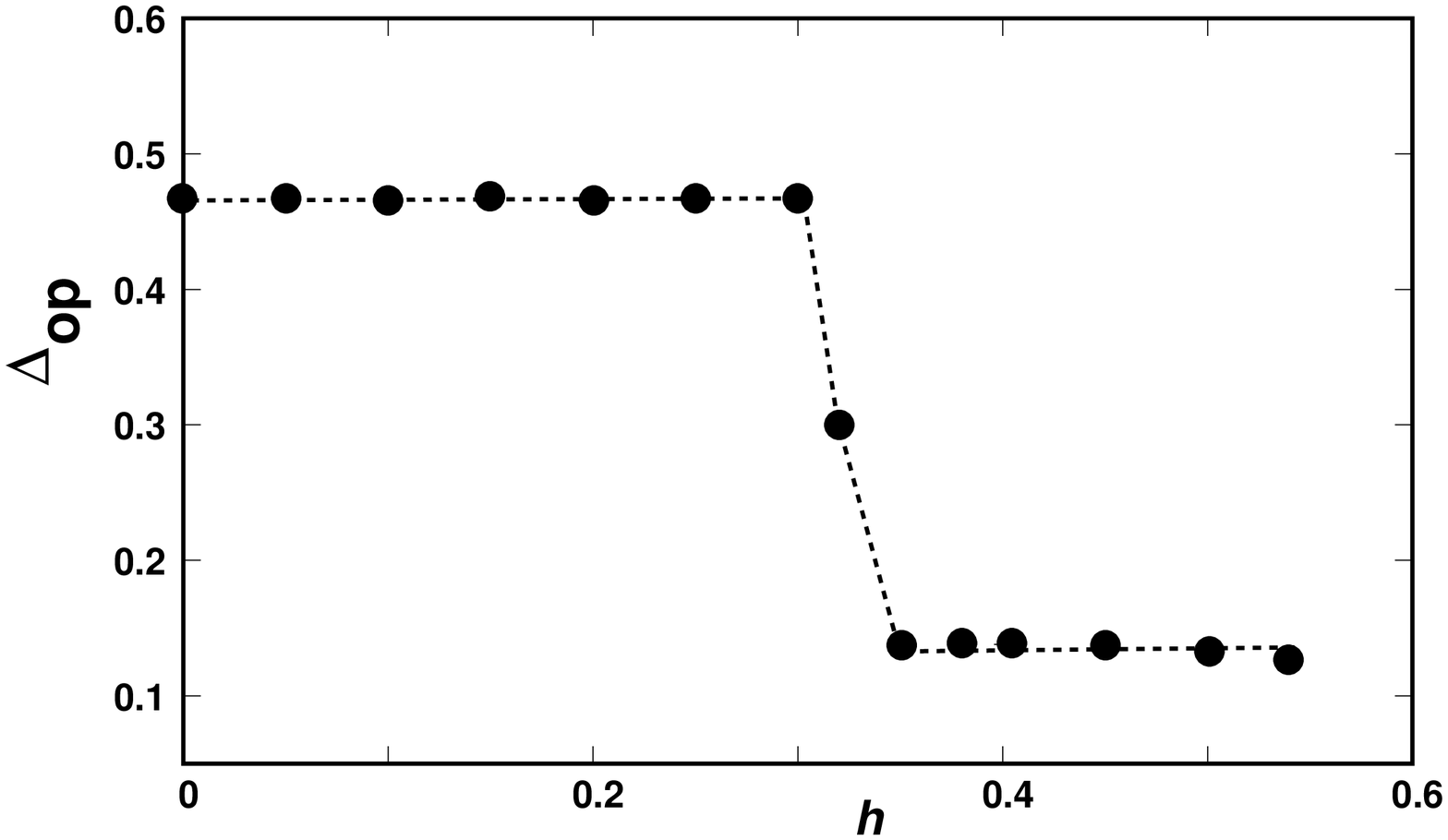}
\caption{The ODLRO order parameter, $\Delta_{OP}$  (in units of $t$) is shown 
 as a function of magnetic field, $h$. The dotted line is a guide to the eye and used in some of the subsequent plots.} 
\end{figure}
 
Next, we compute pair-pair ($ C_{ij}$), density-density ($ K^{\sigma \sigma^{'}}_{ij}$)  and
 current-current ($\Lambda^{\alpha \beta}_{ij}$) correlation  functions, the utility of which 
in experiments are illustrated in literature\cite{Aoki,Koponen,Scalapino}.
 All these quantities point towards
a spatially homogeneous (translationally invariant) ground state for the BCS phase, while the 
translational invariance is broken in the FFLO phase resulting in spatial inhomogeneities
of these correlation functions.
 The correlation functions mentioned above need individual 
attention. 

\begin{figure}[!htb]
\includegraphics[width=0.4\textwidth,height=1.8in]{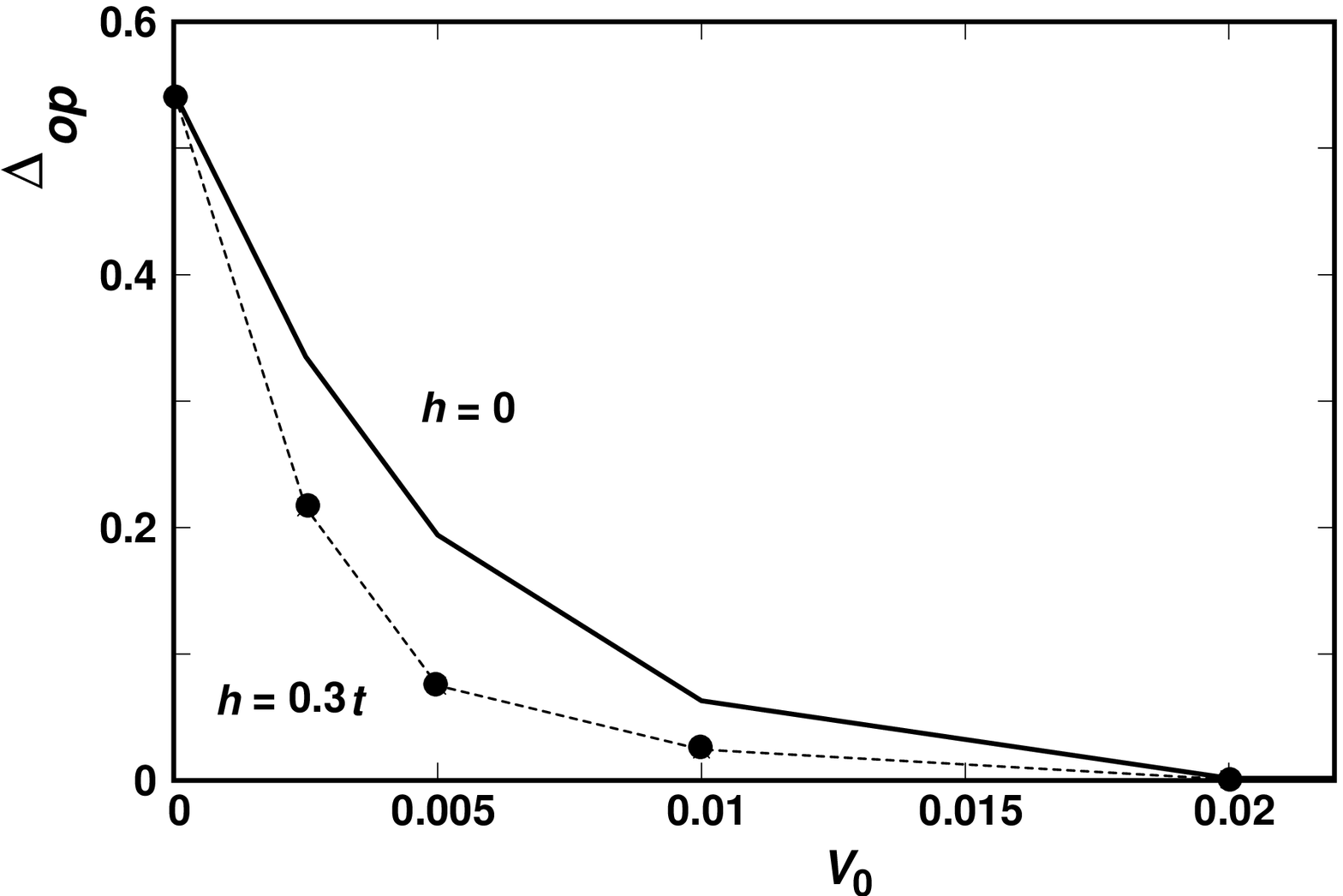}
\caption{The ODLRO order parameter, $\Delta_{OP}$  (in units of $t$) is shown
 as a function of trapping potential, $V_{0}$ for $h=0$ (bold line) and $h=0.3t$ (points and dashed line). 
} 
\end{figure}

We begin with the pair-pair correlation function, $C_{ij}$ (defined in the previous section),
the real space scan of which taken along the length of the lattice (along the direction
in which $\Delta_{i}$ modulates), from one end to another is presented
in Fig.~6(a) and (b). $C_{ij}$ in the  BCS phase ($h=0.1t$) and without an external confinement,
 is higher (approximately ten times) than the value in the FFLO phase ($h=0.5t$). This supports
weakening of the superconducting correlations with  increasing  magnetic field as also is evident from
Figs.~ 2(a) and (c). The behaviour
 of $C_{ij}$ in the presence of trap are interesting in the following sense. It shows
a maximum value at trap center and falls off in a near symmetric fashion 
away from the trap center. It also indicates towards higher correlations in BCS ($h=0$)
 which are considerably reduced (approximately to one-third) in the FFLO
 phase ($h=0.3t$). 
 
 \begin{figure*}[!htb]
\includegraphics[width=0.8\textwidth,height=2.5in]{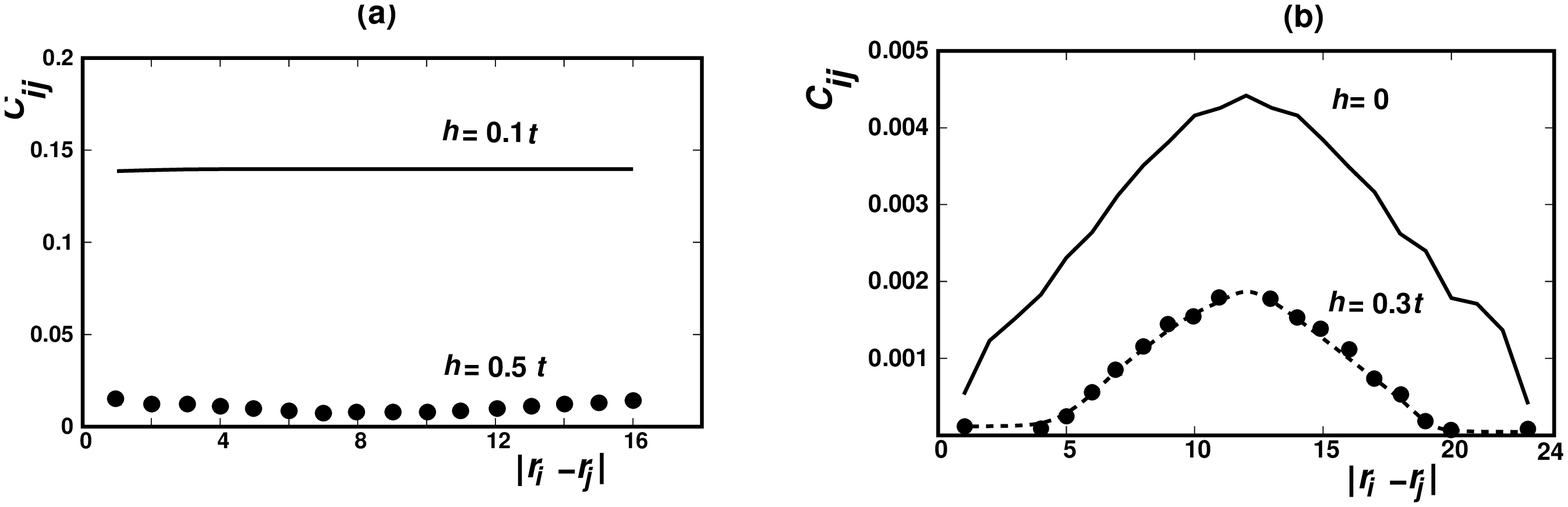}
\caption{  Pair-pair correlation function, $C_{ij}$ is
 shown across the lattice in BCS and FFLO phases without trap (a) and compared with the case 
when the trapping effects are invoked (b).} 
\end{figure*}

The reason behind $C_{ij}$ showing a hump-like behaviour is as follows.
 The probability of finding a pair in the center of the trap is higher
 as the underlying trapping potential is minimum (deep) at the center. However, it drops at the edges
where the trapping potential is  high (shallow). Additionally, $C_{ij}$ drops significantly
 as the external trapping potential is switched on, suggesting the fact that the 
trapping potential assists an imbalanced superfluid phase to have weakened pair-pair correlations.

\begin{figure*}[!htb]
\includegraphics[width=0.8\textwidth,height=2.5in]{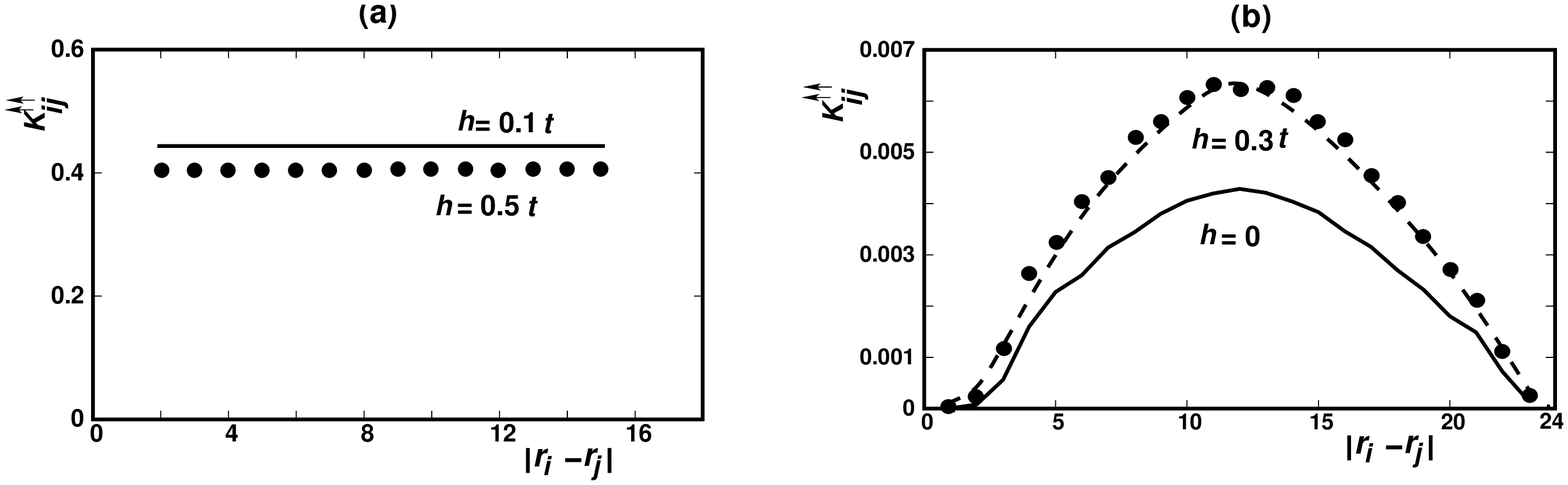}
\caption{  Density-density  correlation function for $\sigma = \sigma^{'} = \uparrow$ {\it{i.e.}}
 $ K^{\uparrow \uparrow}_{ij}$ is
 shown across the lattice in BCS and FFLO phases without trap (a) and with trap (b).  } 
\end{figure*}

Hence we discuss  density-density  correlation function, $ K^{\sigma \sigma^{'}}_{ij}$.
In  absence of the trap, there is no significant
difference between the behaviour of $ K^{\sigma \sigma^{'}}_{ij}$ in the BCS and FFLO phases
(see Fig.~7(a)) with both  remaining constant throughout the lattice.
However, the  correlations show a hump-like 
  feature as the trapping potential is turned on (Fig.~7(b)). Though the shape of the
plot is similar to that of $C_{ij}$, the qualitative behaviour in both the phases 
is reversed {\it{i.e.}} $ K^{\sigma \sigma^{'}}_{ij}$ for the FFLO phase
 ($h=0.3t$) is higher than the BCS phase ($h=0$). It may be noted that we have considered
all $\sigma, \sigma^{'} = \uparrow, \downarrow$. However the results are only shown for 
$\sigma = \sigma^{'} = \uparrow$. Other choices yield qualitatively same results.
 A possible explanation for the rise in correlations between 
different species ($\uparrow$ and $\downarrow$)
 of atoms for the FFLO phase can be explained in terms of the increase in number
 of unpaired particles with the increase in the magnetic field. Thus the correlations
are higher in the FFLO phase (with more number of
unpaired carriers) as compared to the BCS phase where all carriers are paired. Further,
a comparison of  $ K^{\sigma \sigma^{'}}_{ij}$ in the presence of trap to
the one without trap shows remarkable reduction in the  amplitude 
in the presence of trap. These correlations are particularly useful in {\it{time of flight}} experiments
 in which the trapping potential is suddenly switched 
 off, resulting in free expansion of the atomic gas and thus enable realization of the
velocity or momentum profile\cite{Koponen}. These
 techniques have been used to detect the nature of the condensate 
{\it{e.g.}} the superfluidity of the atoms constituting the condensate etc\cite{Altman}.

Finally, we analyze the behaviour of the paramagnetic current-current correlations (with $\alpha = \beta
= x$), {\it{i.e.}} $\Lambda^{xx}_{ij}$ in an environment of harmonic confinement. $\Lambda^{xx}_{ij}$ 
rapidly decays to zero
in the BCS phase, while it modulates across the lattice and remains finite in the FFLO phase,
irrespective of the value of the trapping potential as shown in Fig.~8(a) and (b).
The modulation in real space does not have a uniform profile as $\Delta_{i}$ (Fig.~2(c)), 
possibly because of the involvement of more than one momentum vector in $\Lambda^{xx}_{ij}$.
Nevertheless, we still have the response of the system to an external perturbation, to be more
in the FFLO phase than that in the BCS phase and thus in turn reiterates the presence
of stronger superconducting correlations in the homogeneous BCS phase.
We have recorded higher values for $\Lambda^{xx}_{ij}$ for the case where no trap effects are 
included than those in presence of the trap.
The likely reason must be the reduced mobility of the charge carriers, thereby causing 
weakened current vectors in presence of the trap due to localization effects.

\begin{figure*}[!htb]
\includegraphics[width=0.8\textwidth,height=2.5in]{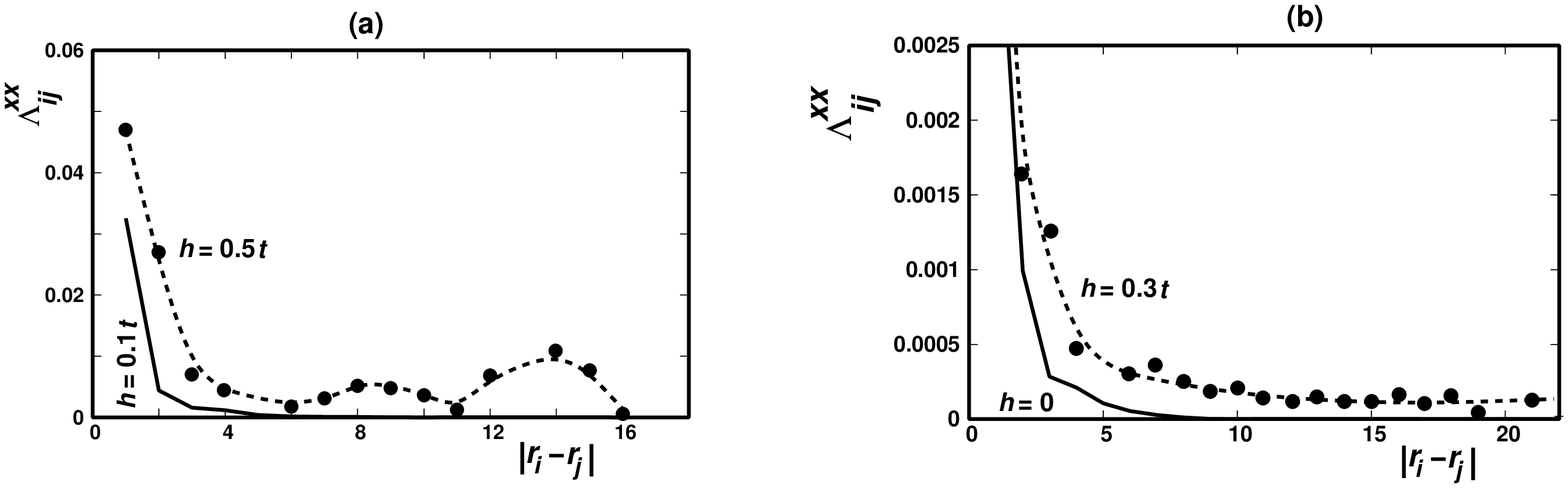}
\caption{  Current-current correlation for $\alpha = \beta = x$ {\it{i.e.}}  $\Lambda^{xx}_{ij}$ is
 shown across the lattice in BCS and FFLO phases without trap (a) and with trap (b).  } 
\end{figure*}

In the following discussion, we describe the scenario corresponding to strong confining 
effects. As the trap is made deeper, in order to keep the chemical potential 
fixed (needed to obtain a population imbalanced state and FFLO phase to be realized), 
the particle density is lowered. So the system looses particles and heads 
towards an {\it{empty}} state with only a few particles remaining at the core of the
potential as trapping effects are made stronger (table I).

\begin{table}
\unitlength1cm
\caption{The total density of particles, $\langle n \rangle (= \sum_{i,\sigma} \langle n_{i\sigma} \rangle $) 
is shown as a function
of increasing trapping strength, $V_{0}$ (in units of $t$). The density drops significantly as the trapping
strength is made stronger leading to an {\it{empty}} system at large values of
$V_{0}$.}
\begin{center}
\begin{minipage}[t]{8.5cm}
\begin{tabular}{|c|c|c|}
\hline 
  & \\
~~Trapping strength~~~~~ & ~~~ Total density ~~~~     \\ 
~~($V_{0}$) ~~~~~    & ~~~($\langle n \rangle$ )~~ \\ \hline

 ~~0.00  ~~ &  0.6361   \\
 ~~0.016 ~~ &  0.2135    \\
 ~~0.05  ~~ &  0.0659    \\
 ~~0.10 ~~ &   0.0313   \\  
 ~~0.20 ~~ &   0.0156    \\ \hline
\end{tabular}
\end{minipage}
\end{center}
\end{table}

In order to obtain a detailed understanding of what happens at larger trap depths, we 
compute certain physical properties that characterize the state. The gap parameter
 ($\Delta_{i}$) vanishes for this scantily populated system (not shown here). The
spectral gap, $E_{gap}$ which is defined as the difference in energy
between the ground state and lowest lying excited state of the BdG spectrum
as obtained by numerically solving Eq.~(\ref{eq:matrix}), shows an increase
with increasing $V_{0}$. There is possibly a very shallow minimum that exists
around $V_{0} \sim 0.016t$ (Fig.~9(a)) as seen in our numerical computation, however
this feature is not clear from  Fig.~9(a). Since an increasing $E_{gap}$ suggests the onset
of an insulating behaviour\cite{Ghosal}, we explored the average kinetic energy 
of the particles (Fig.~9(b)) as a measure of their mobilities and find a rapid decay as $V_{0}$
is increased.

\begin{figure*}[!htb]
\includegraphics[width=0.8\textwidth,height=2.5in]{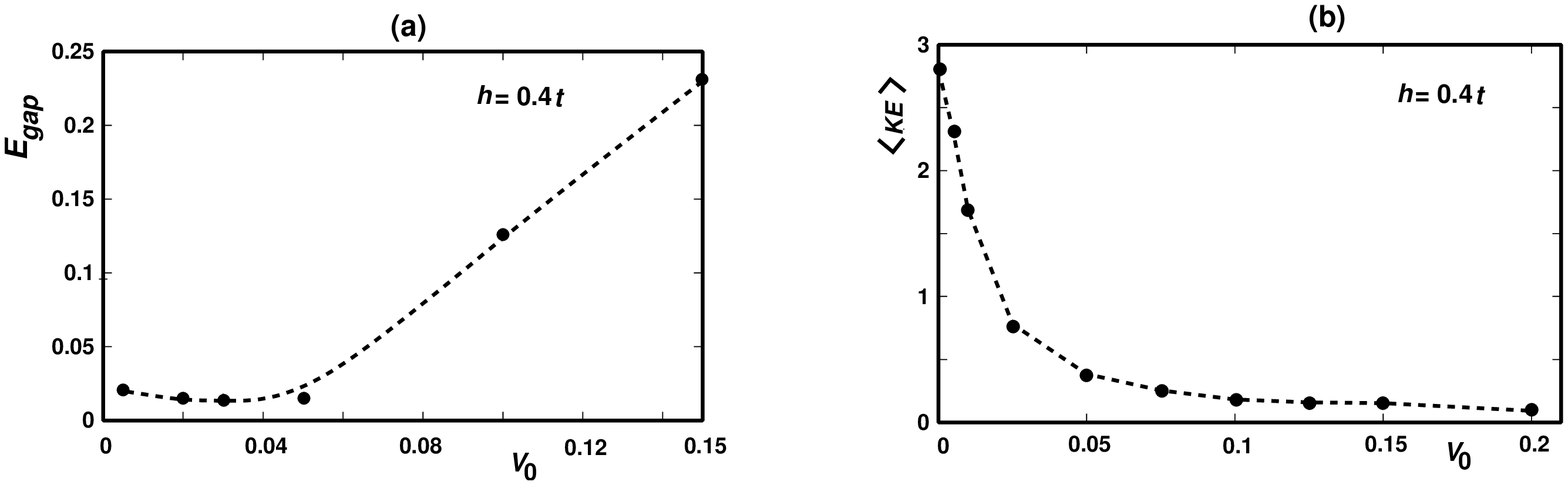}
\caption{(a) The spectral gap $E_{gap}$  is shown as a function of increasing trapping potential $V_{0}$.
(b) The  average kinetic energy of the charge carriers, $\langle KE \rangle$ 
is plotted as a function of $V_{0}$. Both plots are presented at $h = 0.4t$. } 
\end{figure*}

All these results put together may suggest of the onset of an insulating phase at large trap depths,
however when taken together with the fact that particles are gradually {\it{leaking out}} (see table I),
they appear to be artifacts of emptying of the system.

\section{Conclusions}

In summary, we have explored the ground state of a population imbalanced
Fermi gas on a two dimensional lattice in presence of a harmonic
confinement. We adopted a weak coupling $s$-wave superconductor described
by an attractive Hubbard model in presence of imbalanced spin species
induced by a Zeeman field. The ground state is inhomogeneous for a range of
magnetic field strengths which is characterized by a modulating order
parameter, {\it{i.e.}} the FFLO phase. The exotic FFLO phase is obtained both
in presence and absence of the trap.
 
To aid our understanding of the trapping effects on the system, we computed various correlation functions {\it{e.g.}}
  off-diagonal long range order, pair-pair, density-density and 
current-current correlations in real space and provided adequate 
comparisons with the scenario  when the confining potential is absent. The results are indicative
of weakened correlations as population imbalance is increased and confining effects are invoked.

 Larger confining effects induce loss of particles from the system (chemical potential being fixed to its
 original value). At large values of $V_{0}$, the system becomes almost {\it{empty}} of particles. The
 physical properties of this {\it{empty}} system behaves similar to that of an insulator. However
 such signatures should {\it{not}} be considered indicative of a transition from a 
 superfluid to an insulating phase as they are artifacts of a near {\it{empty}} system.

\begin{acknowledgments}
One of us (P.D.) acknowledges C. Y. Kadolkar for helpful discussions.
We thank CSIR and DST, India for financial support under the
Grants  - F.No:09/731(0048)/2007-EMR-I, No.03(1097)/07/EMR-II and SR/S2/CMP-23/2009. 
\end{acknowledgments}


\begin{thebibliography}{99}
%
\bibitem{Anderson} M. H. Anderson, J. R. Ensher, M. R. Matthews, C. E. Wieman, 
and E. A. Cornell, Science {\bf{269}}, 198 (1995).
%
\bibitem{Han} D. J. Han, R. H. Wynar, P. Courteille, and D. J. Heinzen,
Phys. Rev. A {\bf{57}}, R4114 (1998).
%
\bibitem{Ernst} U. Ernst, A. Marte, F. Schreck, J. Schuster, and G. Rempe,
 Europhys. Lett. {\bf{41}}, 1 (1998).
 %
 \bibitem{Esslinger}  T. Esslinger, I. Bloch, and T. W. Hansch, Phys. Rev. A
{\bf{58}}, R2664 (1998).
%
\bibitem{Davis} K. B. Davis, M. -O. Mewes, M. R. Andrews, N. J. van Druten, D. S. Durfee, D. M. Kurn, and W. Ketterle, Phys. Rev. Lett. {\bf{75}}, 3969 (1995).
 %
\bibitem{McNamara} J. M. McNamara, T. Jeltes, A. S. Tychkov, W. Hogervorst, and W. Vassen, Phys. Rev. Lett. {\bf {97}}, 080404 (2006).
%
\bibitem{Fukuhara} T. Fukuhara, Y. Takasu, M. Kumakura, and Y. Takahashi, Phys. Rev. Lett. {\bf {98}}, 030401
(2007); T. Fukuhara, S. Sugawa, and Y. Takahashi, Phys. Rev. A {\bf {76}}, 051604(R) (2007). 
%
\bibitem{Chen} Y. Chen,  Z. D. Wang, F. C. Zhang, and C. S. Ting, Phys. Rev. B {\bf{79}}, 054512 (2009).
%
\bibitem{Iskin} M. Iskin and C. J. Williams, Phys. Rev. A {\bf{78}}, 011603(R) (2008).
%
\bibitem{Loh} Y. L. Loh and N. Trivedi, Phys. Rev. Lett. {\bf {104}}, 165302 (2010).
%
\bibitem{Chandrasekhar} B. S. Chandrasekhar, Appl. Phys. Lett. {\bf{1}}, 7
 (1962).
%
\bibitem{Jordens} R. J\"ordens, N. Strohmaier, K. G\"unter, H. Moritz, and T. Esslinger, Nature {\bf{455}}, 204 (2008).
%
\bibitem{Greiner} M. Greiner, O. Mandel, T. Esslinger, T. W. Hansch, and I. Bloch, Nature {\bf{415}}, 39 (2002).
%
\bibitem{Zwerger} Zwerger, J. opt. B {\bf{5}}, 59 (2003).
%
\bibitem{Stoferle} T. St\"oferle, H. Moritz, K. G\"unter, M. K\"ohl, and T. Esslinger, Phys. Rev. Lett {\bf{96}},
030401 (2006).
%
\bibitem{Strohamair} N. Strohmaier, Y. Takasu, K. G\"unter, R. J\"ordens, M. K\"ohl, H. Moritz, and T. Esslinger,  Phys. Rev. Lett {\bf{99}}, 220601 (2007).
 %
\bibitem{Koetsier} A. Koetsier, R. A. Duine, I. Bloch, and H. T. C. Stoof, Phys. Rev. A {\bf{77}}, 023623 (2008).   
 %
\bibitem{Liu1} X. J. Liu, H. Hu, and P. D. Drummond, Phys. Rev. A {\bf{76}}, 043605 (2007).
%
\bibitem{Liu2} X. J. Liu, H. Hu, and P. D. Drummond, Phys. Rev. A {\bf{78}}, 023601 (2008).
%
\bibitem{Qinghong} Q. Cui and K. Yang, Phys. Rev. B {\bf{78}}, 054501 (2008).
%
\bibitem{Waldram} J. R. Waldram, {\it{Superconductivity~ of~ Metals~ and~ Cuprates}} 
(Institute of Physics Publishing, London, 1996).
%
\bibitem{Wang} Q.  Wang,   H. -Y. Chen, C. -R. Hu, and C. S. Ting, 
Phys. Rev. Lett. {\bf{96}}, 117006 (2006); Q. Cui and  K. Yang, Phys. Rev. B
{\bf{78}}, 054501 (2008).
%
\bibitem{Aoki} H. Aoki and K. Kuroki, Phys. Rev. B {\bf{42}}, 2125 (1990).
%
\bibitem{Koponen} T. K. Koponen, T. Pannanen, J. -P. Martikainen, M. R. Bakhtiari, and P. T\"orm\"a,
New Journal of Physics {\bf{10}}, 045014 (2008).
%
\bibitem{Scalapino} D. J. Scalapino, S. R. White, and S. Zhang, Phys. Rev. B {\bf{47}}, 7995 (1993).
%
\bibitem{Altman} E. Altman, E. Demler, and M. D. Lukin Phys. Rev. A {\bf{70}}, 013603 (2004).   
%
\bibitem{Ghosal} A. Ghosal, M. Randeria, and N. Trivedi, Phys. Rev. B {\bf{65}}, 014501 (2001).


\end{thebibliography}
\end{document}